# On-Package Memory with Universal Chiplet Interconnect Express (UCIe): A Low Power, High Bandwidth, Low Latency and Low Cost Approach

Debendra Das Sharma[1], Swadesh Choudhary[1], Peter Onufryk[1], and Rob Pelt[2]

[1]Intel Corporation and  [2]AMD Corporation

***Abstract*** – Emerging computing applications such as Artificial Intelligence (AI) are facing a memory wall with existing on-package memory solutions that are unable to meet the power-efficient bandwidth demands. We propose to enhance UCIe with memory semantics to deliver power-efficient bandwidth and cost-effective on-package memory solutions applicable across the entire computing continuum. We propose approaches by reusing existing LPDDR6 and HBM memory through a logic die that connects to the SoC using UCIe. We also propose an approach where the DRAM die natively supports UCIe instead of the LPDDR6 bus interface. Our approaches result in significantly higher bandwidth density (up to 10x), lower latency (up to 3x), lower power (up to 3x), and lower cost compared to existing HBM4 and LPDDR on-package memory solutions.

## I. Introduction

For decades, we've been in a virtuous cycle of innovation, with major advances in computing capabilities, largely driven by breakthroughs in process and packaging technologies enabling a wide range of applications. Some applications such as Artificial Intelligence require an annual ten-fold increase in compute capability within a fixed power profile. Memory has been the bottleneck from a bandwidth and capacity point of view, resulting in inefficient use of computing resources, negatively impacting both power consumption and infrastructure costs[1,2,3]. Computing, which already consumes a significant share of global energy[4], is facing an unsustainable future unless more power-efficient solutions are developed.

The prevalent approach in commercial systems to deliver power-efficient and cost-effective performance is heterogeneous integration of memory, compute, and communication chiplets on the same package. For example, smartphones, tablets, laptops, and PCs integrate Low Power Double Data Rate (LPDDR) DRAM on-package to reduce power consumption, board area, and cost. Examples include Apple's M-series[5,6] (Figure 1a), Intel's Lunar Lake AI PC[7], and Qualcomm's Snapdragon processors[8]. In contrast, high-performance computing (HPC), AI, and server applications commonly use the more expensive (5-10x per bit compared to LPDDR[9-11]) on-package High Bandwidth Memory (HBM) to deliver higher bandwidth (e.g., 20x bandwidth of LPDDR for same capacity[12-14]). Notable examples include AMD Instinct MI300 Series[15], AWS Trainium[16], Google TPU[17], Intel's Xeon Max CPU[18] (Figure 1b), Data Center GPU Max[19], Altera FPGA[20], Nvidia Hopper H100 GPU[21] (Figure 1c), and AMD FPGA[22].

Memory interfaces such as LPDDR and HBM continue to rely on parallel, bi-directional buses, with separate buses for command and data, each operating at different frequencies, to be compatible with DRAM processes. This approach leads to low bandwidth and pin/package/board inefficiencies due to much lower frequencies compared to on-package interconnects like UCIe[23-25] and off-package interconnects such as Compute Express Link (CXL)[26]. Next generation LPDDR6[27] and HBM4[28] are following the same approach, which will result in continued memory bottlenecks in a power constrained world. The quest for power-efficient and cost-effective bandwidth scaling led the industry to shift from bi-directional multi-drop bus-based interconnects to serial point-to-point interconnects for I/O over the past two decades. Key examples of this transformation include the transition from the PCI bus to PCI-Express, the switch from front-side bus to links for scale-up cache coherency interconnects, and more recently, the adoption of CXL for memory attachment at the platform level. We believe on-package memory needs to undergo a similar revolution by adopting a unidirectional, serialized, point-to-point interconnect like UCIe, as proposed in this paper. This will result in higher bandwidth and lower power (peak, idle, and power consumption proportionate to bandwidth consumption due to the low latency to turn on/off circuits), despite the challenges associated with asymmetric memory accesses with a unidirectional link, compared to existing LPDDR and HBM approaches, as demonstrated in this paper.

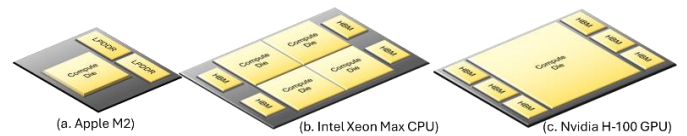

*Figure 1: Compute die(s) and Memory (LPDDR and HBM) connected on-package in various commercial offerings*

## II. UCIe Specification: An Overview

UCIe 1.0 and 2.0 specifications[23-25,29-30], define two types of packaging for planar interconnect: standard or 2D (UCIe-S) and advanced or 2.5D (UCIe-A), as shown in Figure 2a. The standard package is used for cost-effective performance. The advanced packaging is used for power-efficient performance but with higher packaging cost.

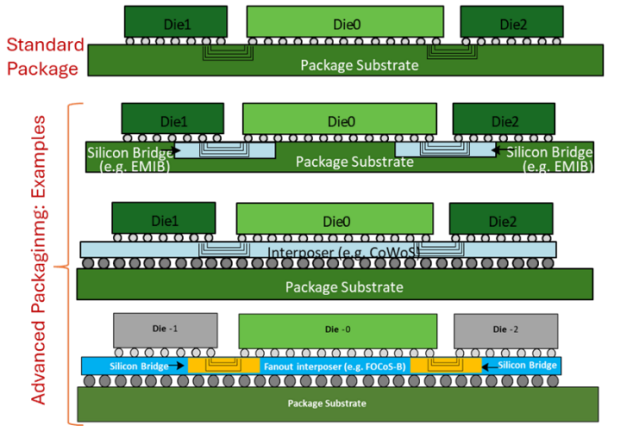

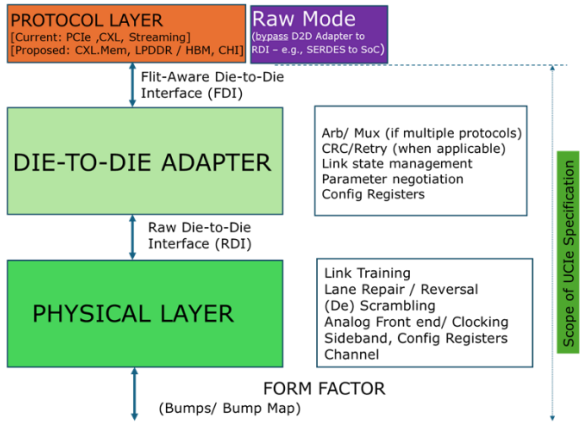

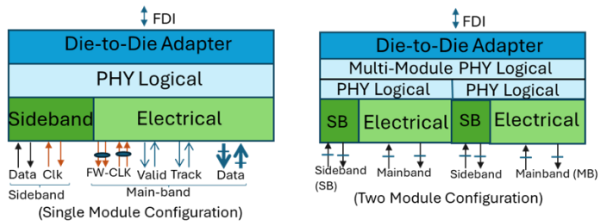

*Figure 2: Layered and scalable approach of UCIe (2D and 2.5D)*

UCIe is a layered protocol, as shown in Figure 2b. The physical layer (PHY) specifies multiple data rates, bump map, channel model, electrical signaling, clocking, link training, sideband, circuit architecture, etc. The basic unit is a module (Figure 2c) which comprises of N single-ended, unidirectional, full-duplex 'Data' lanes (N=16 for UCIe-S and N=64 for UCIe-A), one single-ended lane for valid, one lane for tracking, a differential forwarded clock per direction for the main-band. The sideband consists of 2 single-ended lanes (one data and one 800 MHz forwarded-clock) per direction. The sideband interface is used for link training, register access, and diagnostics. Multiple modules (1, 2, or 4) can be aggregated to deliver more performance per link (Figure 2c). The D2D adapter is responsible for reliable delivery of data through its cyclic redundancy check (CRC) and link level retry mechanism using a 256-Byte flow-control unit (Flit). Table 1 summarizes the key metrics of UCIe. The protocol layer can be any protocol, including the memory protocols we propose in this paper, using the Flit Mode or the Raw Mode which bypasses the D2D adapter.

*Table 1: Key Metrics of UCIe*

| Metrics | UCIe-2D[1] | UCIe-2.5D[1] | UCIe-3D[1] |
|---|---|---|---|
| Data Rate (GT/s) | 4, 8, 12, 16, 24, 32 | 4, 8, 12, 16, 24, 32 | <= 4G |
| Width (per direction) | 16 | 64 | 80 |
| Bump Pitch (μm) | 100-130 | 25-55 | <=1 - 9 |
| Channel Reach | 25 mm | 2 mm | ~0 mm (Hybrid Bonding) |
| B/W Shoreline (GB/s/mm) | 28-224 | 165-1317 | N/A (areal only) |
| B/W Density (GB/s/mm$^2$) | 22 – 125 | 188-1350 | 4000 (9μ) – 300,000 (1μ) |
| | B/W corresponds to frequency. UCIe 2D @ 110 μm; 2.5D @ 45 μm | | At 4G |
| Power Efficiency (pJ/b) | 0.5 (<=16G) / 0.6 (>16G) | 0.25 (<=16G) / 0.3 (>16G) | 0.05 (9μ) - 0.01 (1 μ) |
| Dynamic Power Savings | <1ns entry/exit with 85%+ power savings | | |
| Latency (round-trip) | 2ns | | < 1ns |

[ [1]: From UCIe 1.0 and 2.0 Specifications – we are bump-limited with UCIe[23-25,29,30] ]

### III. OUR PROPOSED APPROACHES FOR UCIe MEMORY

We propose UCIe-Memory; architecture and interconnect to connect compute (or communication) chiplet (referred as System-on-a-Chip, SoC, in this paper) with on-package memory, using UCIe electrical signaling, as summarized in Figure 3.

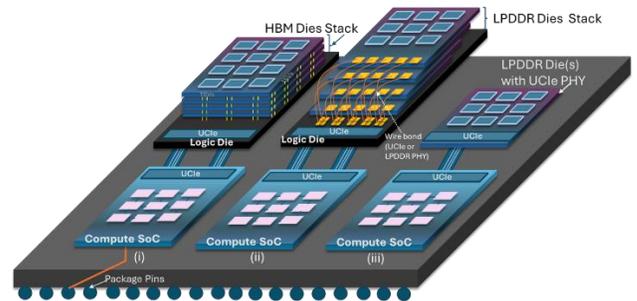

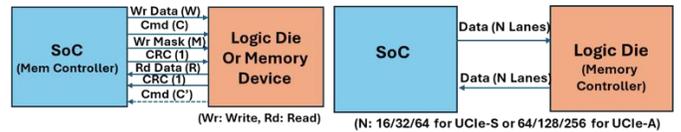

*Figure 3: Proposed Architectures and Protocol for UCIe-Memory*

Our proposed approaches fall into two broad categories:

1. The SoC connects to a logic die using UCIe (like the HBM approach). The logic die connects to either a native HBM 3D memory stack, as shown in Figure 3 a,i, or a stack of native LPDDR dies, using wire-bonding on-package, as shown in Figure 3 a,ii. The approaches in this category do not require any changes to DRAM devices and likely to be deployed first. We are simply replacing the front-end of the logic-die from HBM bi-directional bus with UCIe link. The memory controller can reside in two places with this method: (i) The memory controller resides in the SoC, as shown in Figure 3b. The logic die takes care of the DRAM signaling, access timing, and scheduling of events like refresh, like the existing approach with HBM. (ii) The memory controller resides in the logic die, as shown in Figure 3c. The SoC uses higher-level, unordered, and memory-technology independent memory read-write semantics, like CXL.Memory, to access the memory hosted by the logic die. All the approaches proposed below (A through E) fall into this category.

2. The memory die (e.g., LPDDR6) can directly connect to a SoC using UCIe PHY, as shown in Figure 3.a.iii. The memory controller resides in the SoC and takes care of DRAM access and timing, as shown in Figure 3b. This method needs changes to the DRAM device and likely take longer than the first approach for deployment. However, this will be useful for segments such as the hand-held computing devices. One of the options in approach A below, belongs to this category. It should be noted that this category can be used simultaneously with category 1 (where the SoC to logic die is category 1 and logic die to LPDDR is category 2).

UCIe-Memory interconnect can have multiple options, while using UCIe PHY. Some options rely on the existing symmetric UCIe, e.g., 16 or 64 data lanes in each direction per cluster, as shown in Figure 3c. We also propose enhancements for 'Asymmetric UCIe', where each direction has different width, for direct connect to memory die (Figure 3b, e.g., commands are only issued from SoC to memory). We also propose to enhance UCIe PHY to operate at a multiple of the underlying DRAM frequency or run the DRAM at a fractional UCIe frequency. This results in better bandwidth efficiency and avoids any asynchronous clock domain crossings. For example, if DRAM runs at 8333 MT/s, it can run at 8 GT/s and UCIe can run at 8/16/32 GT/s or if the DRAM runs at 10GT/s, the UCIe link can run at 10/20/40 GT/s.

The UCIe sideband interface can be used to access configuration registers, report error/ events/ performance monitors, and signal interrupts as needed in all the proposed approaches below. These are not performance-critical; the sideband bandwidth of 800 Mb/s/direction is enough to meet these needs while simplifying the main data path by removing CXL.io[23-25].

A. LPDDR6 protocol mapping on Asymmetric UCIe: Here the memory controller resides in the SoC. We propose two options. In the first option, the memory controller is in the logic die and the SoC to logic die connection is UCIe, as shown in Figure 4a. Here, the module size is 74 data Lanes (vs multiples of 32 with existing UCIe-S/A). Here the read to write bandwidth ratio is at 3:2, to get closer to a x32 UCIe bump map with the addition of another row of signals.

In the second option, the LPDDR6 die is modified to support UCIe PHY natively, as shown in Figure 4b. Here the UCIe PHY runs all wires at the same frequency as the data (DQ) in the DRAM to be DRAM-process friendly, though it can easily run at 2x or 4x of the DQ frequency to be bump-efficient. The data to each device array is 12 bits wide in each direction, 2 command bits go to both device arrays (but can be modified to 4 bits if each device array can support simultaneous read/ write to different banks to deliver higher bandwidth). The UCIe interface is optimized for a read to write bandwidth ratio of 2:1 (hence 24 data from LP6 die to SoC vs 12 write data in the other direction), reflecting the predominant usage models (even a write involves a read followed by a write, though write-only happens with things such as logs which are read only on error conditions). For 100% writes, this 2:1 approach yields half the bandwidth of native LPDDR6 with 24 bi-directional wires, assuming the same frequency on UCIe (though UCIe is capable of up to 4x data rate of LPDDR6). For 100% reads our approach with UCIe has the same bandwidth as LPDDR6, and higher bandwidth for read-write mix, assuming the same operating frequency as LPDDR6. Thus, the UCIe module size is 43 or 45 data lanes and the UCIe PHY can support parallel read/ write to each device array simultaneously.

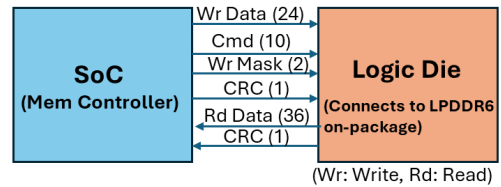

(a) On-Package LPDDR6 through Logic Die

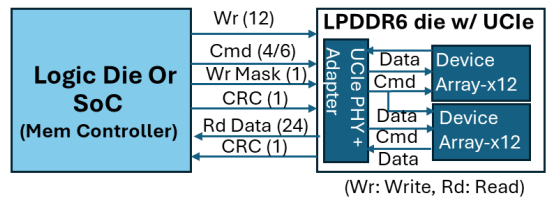

(b) LPDDR6 die with native UCIe PHY

| Signal Name | Direction | Frequency | Count |
|---|---|---|---|
| Command / Address (CA) | SoC->Mem | 1/2 | 8 |
| Chip Select (CS) | SoC->Mem | 1/2 | 2 |
| Wr Clk (wclk_t, wclk_c) | SoC->Mem | 1/2 | 4 |
| Rd Clk (RDQS_t, RDQS_c) | Mem->SoC | 1/2 | 4 |
| Clock (ck_t, ck_c) | SoC->Mem | 1/4 | 4 |
| Data | Bi-directional | 1 | 24 |
| Total (46) | | | 46 |

(c) Signal List for 2 sub-channel x24 LPDDR6

| Signal Name | SoC-> Mem | Mem -> SoC |
|---|---|---|
| Command/ Address | 4 | 0 |
| Data | 12 | 24 |
| Wr Mask | 1 | 0 |
| CRC | 1 | 1 |
| UCIe Data Total | 18 | 25 |
| UCIe Clock, Track, Valid | 2, 1, 1 | 2, 1,1 |
| Total = 51 | 22 | 29 |
| 64B transfer (UI) | 48 | 24 |

(d) LPDDR6 signal mapping on UCIe

*Figure 4: LPDDR6 protocol on Asymmetric UCIe with signal mapping*

The native LPDDR6 signal list is shown in Figure 4c. Its mapping to UCIe-Memory is demonstrated in Figure 4d. The LPDDR6 protocol and timings are maintained as-is in our proposed mapping. An example pipelined timing diagram of efficient conversion between UCIe command and data to/from 4 LPDDR6 dies, keeping the LPDDR6 signal and timing intact, is provided in the appendix.

B. HBM3/4 protocol mapping on Asymmetric UCIe: Here the memory controller resides in the SoC, which connects to a logic die using UCIe; the logic die in turn connects to the stacked HBM memory (Figure 3,a,i). The proposed UCIe module size is 138 data Lanes (Figure 5a) with a read vs write bandwidth ratio of 2:1. The HBM 3D stack runs at 1-2 GT/s, a fraction of UCIe frequency. Details of the signal mapping appears in Figure 5b, preserving the same signal list, protocol, and timings as HBM.

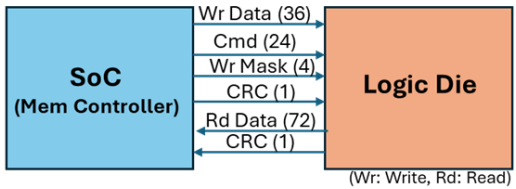

(a. HBM3/4 protocol on UCIe)

| Signal Name | SoC->Logic | Logic->SoC |
|---|---|---|
| Command | 24 | 0 |
| DRAM Data, Wr Mask | 36, 4 | 72, 0 |
| CRC | 1 | 1 |
| Clk,Track,Valid | 2,1,1 | 2,1,1 |
| Total (Data) | 69 (65) | 77 (73) |
| Cache transfer (UI) | 16 | 8 |

(b. HBM3/4 Signal Mapping on UCIe)

*Figure 5: HBM3/4 protocol on Asymmetric UCIe with signal mapping*

| | 0 | 1 | 2 | 3 | 4 | 5 | 6 | 7 | 8 | 9 | 10 | 11 | 12 | 13 | 14 | 15 |
|---|---|---|---|---|---|---|---|---|---|---|---|---|---|---|---|---|
| Byte 0 | LinkHdr0,1 | | | | | | G0 Bytes 0 through 13 | | | | | | | | | |
| Byte 16 | G0 Bytes 14 through 19 | | | | | | G1 Bytes 0 through 9 | | | | | | | | | |
| Byte 32 | G1 Bytes 10 through 19 | | | | | | | G2 Bytes 0 through 5 | | | | | | | | |
| Byte 48 | G2 Bytes 6 through 19 | | | | | | | | | | | | | | ProtHdr0,1 | |
| Byte 64 | ProtHdr2,3 | | G3 Bytes 0 through 13 | | | | | | | | | | | | | |
| Byte 80 | G3 Bytes 14 through 19 | | | | | | G4 Bytes 0 through 9 | | | | | | | | | |
| Byte 96 | G4 Bytes 10 through 19 | | | | | | | G5 Bytes 0 through 5 | | | | | | | | |
| Byte 112 | G5 Bytes 6 through 19 | | | | | | | | | | | | | | LinkHdr2,3 | |
| Byte 128 | ProtHdr4,5 | | G6 Bytes 0 through 13 | | | | | | | | | | | | | |
| Byte 144 | G6 Bytes 14 through 19 | | | | | | G7 Bytes 0 through 9 | | | | | | | | | |
| Byte 160 | G7 Bytes 10 through 19 | | | | | | | G8 Bytes 0 through 5 | | | | | | | | |
| Byte 176 | G8 Bytes 6 through 19 | | | | | | | | | | | | | | ProtHdr6,7 | |
| Byte 192 | ProtHdr8,9 | | G9 Bytes 0 through 13 | | | | | | | | | | | | | |
| Byte 208 | G9 Bytes 14 through 19 | | | | | | G10 Bytes 0 through 9 | | | | | | | | | |
| Byte 224 | G10 Bytes 10 through 19 | | | | | | | G11 Bytes 0 through 5 | | | | | | | | |
| Byte 240 | G11 Bytes 6 through 19 | | | | | | | | | | | | | | LinkHdr4,5 | |

*Figure 6: CHI mapping on symmetric UCIe*

C. Coherent Hub Interface (CHI) on Symmetric UCIe: Here, the memory controller resides in the logic die (Figure 3c). The CHI[32] Flit mapping to UCIe released by ARM is demonstrated in Figure 6. In this approach, CHI is mapped to UCIe using the 256B containers[32], which is based on UCIe latency-optimized Flit[23-25,30]. For memory mapping, we use Format-X since it has twelve 20B granules resulting in better efficiency than Format -Y for memory access. The remaining 16B are used for Link and Protocol Headers, comprising of CRC, FEC, Credits, etc. A granule is 20B in size and carries requests or responses along with data where applicable. We assume implementations will use the Write Push optimizations defined in CHI to get better efficiency for writes being sent to the memory controller.

D. CXL.Mem[3] on Symmetric UCIe: Here the memory controller resides in the logic die. The SoC connects to logic die using standard UCIe module (Figure 3c). The logic die connects to an HBM stack (Figure 3,a,i) or an LPDDR6 stack (Figure 3,a,ii).

We map the CXL.Mem protocol[31] to UCIe using the Flit Layout shown in Figure 7. The command layout for the request (read and write in SoC to Memory direction) and responses (data return and completions for writes in Memory to SoC direction) is shown in Table 2, under the two "Unopt" columns. The 2-byte Flit HDR contains information needed for identifying traffic type using a protocol identifier field (e.g., CXL.Mem, CXL.io, NOP Flit, etc.) and reliable Link operation such as sequence number, ack/nak, etc[30,31]. The first 2B of CRC (CRC0) protects the Bytes 0-127 and the second 2B of CRC (CRC1) protects the rest. The Credit field is used for credit management by the transaction layer.

| | 0 | 1 | 2 | 3 | 4 | 5 | 6 | 7 | 8 | 9 | 10 | 11 | 12 | 13 | 14 | 15 |
|---|---|---|---|---|---|---|---|---|---|---|---|---|---|---|---|---|
| Byte 0 | HDR - 2B | | Slot 0 - 14B H-Slot | | | | | | | | | | | | | |
| Byte 16 | Slot 1 - 16B (G-Slot) | | | | | | | | | | | | | | | |
| Byte 32 | Slot 2 - 16B (G-Slot) | | | | | | | | | | | | | | | |
| Byte 48 | Slot 3 - 16B (G-Slot) | | | | | | | | | | | | | | | |
| Byte 64 | Slot 4 - 16B (G-Slot) | | | | | | | | | | | | | | | |
| Byte 80 | Slot 5 - 16B (G-Slot) | | | | | | | | | | | | | | | |
| Byte 96 | Slot 6 - 16B (G-Slot) | | | | | | | | | | | | | | | |
| Byte 112 | Slot 7 - 16B (G-Slot) | | | | | | | | | | | | | | | |
| Byte 128 | Slot 8 - 16B (G Slot) | | | | | | | | | | | | | | | |
| Byte 144 | Slot 9 - 16B (G-Slot) | | | | | | | | | | | | | | | |
| Byte 160 | Slot 10 - 16B (G-Slot) | | | | | | | | | | | | | | | |
| Byte 176 | Slot 11 - 16B (G-Slot) | | | | | | | | | | | | | | | |
| Byte 192 | Slot 12 - 16B (G-Slot) | | | | | | | | | | | | | | | |
| Byte 208 | Slot 13 - 16B (G-Slot) | | | | | | | | | | | | | | | |
| Byte 224 | Slot 14 - 16B (G-Slot) | | | | | | | | | | | | | | | |
| Byte 240 | Reserved - 10B | | | | | | | | | | | Credit-2B | | CRC0-2B | CRC1-2B | |

*Figure 7: CXL.Mem mapped to 256B UCIe Flit (Unoptimized)*

Slots form the basic unit of transfer in the 256-byte Flit which contains one H-Slot and 14 G-Slots; like CXL[31]. Even though the CRC appears at the end, the memory controller can schedule the memory accesses as soon as it receives the command to reduce the latency on reads. If a Flit containing a read command has an error, the read data is ignored. The SoC will retry that Flit again. Writes are not latency sensitive and may be buffered to ensure it passes CRC and other D2D adapter framing checks. Commands (Request/ Response) can be sent in H- or G-slot whereas data (to/from DRAM) is sent only on G-slot. Each SoC->Mem Request is 74 bits and each Mem-> SoC response is 26 bits. Thus, one request or two responses can fit in each slot, following CXL rules. The Flit scheduling mechanism optimizes by packing as many headers as possible into an H-slot and leave as many G-slots for data to drive the best efficiency in the Link.

*Table 2: Command fields for CXL.Mem: Optimized and Unoptimized*

| Field Name | SoC ->Mem Req (Rd/Wr) | | Mem -> SoC Resp (Data, Cmpl) | |
|---|---|---|---|---|
| | Unopt | Opt | Unopt | Opt |
| Cmd | 4 | 3 | 3 | 3 |
| Meta Data | 7 | 4 | 4 | 4 |
| Devload | 0 | 0 | 2 | 0 |
| Tag | 16 | 8 | 16 | 8 |
| Address | 46 | 46 | 0 | 0 |
| Poison | 1 | 1 | 1 | 1 |
| Total | 74 | 62 | 26 | 16 |

|   | 0 1 2 3 | 4 5 6 7 8 9 | 10 11 12 13 | 14 15 |
|---|---|---|---|---|
| Byte 0 | Slot 0 - 16B (G-Slot) | | | |
| Byte 16 | Slot 1 - 16B (G-Slot) | | | |
| Byte 32 | Slot 2 - 16B (G-Slot) | | | |
| Byte 48 | Slot 3 - 16B (G-Slot) | | | |
| Byte 64 | Slot 4 - 16B (G-Slot) | | | |
| Byte 80 | Slot 5 - 16B (G-Slot) | | | |
| Byte 96 | Slot 6 - 16B (G-Slot) | | | |
| Byte 112 | Slot 7 - 16B (G-Slot) | | | |
| Byte 128 | Slot 8 - 16B (G-Slot) | | | |
| Byte 144 | Slot 9 - 16B (G-Slot) | | | |
| Byte 160 | Slot 10 - 16B (G-Slot) | | | |
| Byte 176 | Slot 11 - 16B (G-Slot) | | | |
| Byte 192 | Slot 12 - 16B (G-Slot) | | | |
| Byte 208 | Slot 13 - 16B (G-Slot) | | | |
| Byte 224 | Slot 14 - 16B (G-Slot) | | | |
| Byte 240 | Slot 15 - HS Slot (10B) | | HDR (2B) Credit 2B | CRC (2B) |

*Figure 8: CXL.Mem with optimization mapped to 256B UCIe Flit*

E. CXL.Mem[31] with optimization on Symmetric UCIe: This approach is like our prior approach with optimizations to the Flit (Figure 8). Optimizations to the command for requests to memory and responses to SoC, is shown in Table 2, under the two "Opt" columns.

Our proposed 256-Byte Flit layout with optimization results in 15 G-Slots and one HS-Slot (which will be used for Headers only), resulting in an extra G-Slot per Flit over our prior approach in Figure 7. Here, the HDR and Credit are same as before, but the 2B CRC covers the entire Flit. While the HDR at the end works well for the reliable link operation, the identification of the traffic type may pose a challenge if multiple protocols (beyond CXL.Mem, such as CXL.io) are supported and latency is critical. We propose to overcome that by the receiver parking the protocol identifier of received Flits with NOP after Link (re)training, and the protocol identifier in the HDR bytes identifies the protocol for the next Flit. Since we are targeting on-package memory, we do not need to scale to thousands of nodes with several micro-seconds latency as the native CXL. Hence, several header fields (e.g., Cmd, Addr, Tag) have been reduced (Table 2) to pack more headers in a slot. With command optimization we can fit one request or four responses (vs two) in a 10-byte HS-slot. It is possible to fit two requests in a G-slot (vs one) as an additional optimization, though this is not considered in the performance analysis below.

IV. ANALYSIS AND RESULTS

A. Micro-Architecture

Figure 9 shows the representative micro-architectural diagram in implementations for (symmetric) UCIe at 32GT/s[25,29]. Asymmetric UCIe is similar in latency and proportionate (to bandwidth) gate-count, based on our analysis, but has different widths in the inbound vs outbound direction. It should be noted that other frequencies (not a power of 2) will have 1/16 ratio for the internal clock.

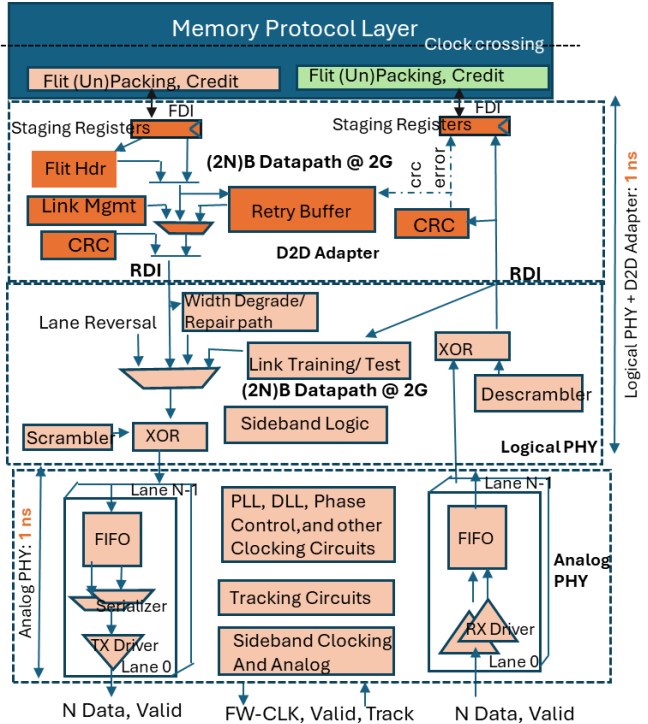

*Figure 9: Micro-architecture of UCIe-Memory at 32GT/s*

The analog PHY has the Tx/Rx drivers along with other circuits for clocking, tracking, and the sideband. There is a FIFO in each direction to adjust to the frequency of the logic domain (1/16th of the forwarded clock frequency) and handle drift. The transmit path uses multiplexers to serialize. Deserialization on Receive side is done through a FIFO. With a 2GHz logic clock, Analog PHY transmit/ receive consumes 0.5 ns each, resulting in a 1ns round-trip.

The Logical PHY has a single level of ex-or for each received/ transmitted data bit for (de)scrambling, using precomputed values from prior cycle. The CRC logic (transmit and receive) has 5 levels of gates. The Flit Hdr generation is done after the arb/mux logic. The rest of the data-path is mux/demux, along with the round-trip latency of 2ns at 2GHz (between FDI and bump), shown in Figure 9. The flit packing and unpacking logic at 2GHz is one clock cycle each, resulting in a round-trip of 3ns from the memory protocol layer.

The equivalent measured latency in our silicon implementation for LPDDR5 is 7.5ns and HBM3 is 6ns with similar results expected in LPDDR6 and HBM4 respectively. Thus, UCIe-Memory outperforms existing approaches from a latency perspective.

*B. Bandwidth Density and Power*

We evaluate the performance of the proposed approaches with that of existing approaches of using LPDDR and HBM on-package memory. Since LPDDR6 and HBM4 specifications are in flight, we use the area and power numbers from our implementation of LPDDR5 and HBM3 to project the different metrics. For our proposed UCIe approaches, we use the power and area number of existing implementations, since we are using the UCIe stack. For the bandwidth density, we use the bump area as the upper bound since UCIe expects designs be bump-limited (unlike HBM and LPDDR); the module fits within the specified bump area, as has been reported through multiple implementations[23,25,26,29,30], including ours.

Here we evaluate are the bandwidth density (both linear and areal) and power efficiency metrics of various approaches, including existing ones like LPDDR and HBM. LPDDR5 and HBM3 implementations report the data bus (DQ) bandwidth transfer in their efficiency calculations. LPDDR6 and HBM4 will follow the same methodology. We use the same methodology for our proposed approaches for uniformity. Thus, CRC, ECC, header, credit, command, and address are considered overhead. Any reserved lanes are considered as overhead for BW efficiency but assumed to be turned off for power calculations. We use multiple representative traffic mixes; represented as $x$ reads and $y$ writes ($xRyW$; $x \geq 0, y \geq 0$ but both cannot be 0) for comparison. The data transferred with $xRyW$ is $512.(x + y)$ bits

$xRyW$ results in $(x + y)$ request headers (from SoC) to memory, $y$ cache line data writes to memory, and $x$ cache line data returns from memory (to SoC). Additionally, there will be $(x + y)$ response headers to SoC when memory controller is on the logic die and CXL.Mem is mapped on UCIe. When the memory controller is in SoC and connected to memory directly (i.e., LPDDR6 and HBM3/4 mapped on UCIe) there are no responses.

We assume that there is no penalty for bus turn-around times for the existing approaches of LPDDR and HBM. We also assume that these bidirectional buses always deliver peak data bandwidth for any traffic mix. We also assume that these are bump-limited (vs circuit limited), even with increased frequencies, even though that is not the case in existing designs. We feel these optimistic projections from existing LPDDR and HBM approaches represent the upper bound and simpler to compare without getting into implementation details of multiple LPDDR and HBM designs that vary across different designs and process technologies.

For LPDDR5, the bump map is 5.8 mm (die edge) x 1.75 mm (depth) for 128 DQ (data) at 9.6 GT/s[12]. For LPDDR6, we assume the same linear and areal density for its 192 DQ operating at 9.6 – 12.8 GT/s, even though there may be a higher ratio of power/gnd to signal pins due to the width and speed increase. Thus, for LPDDR-5 at 9.6 GT/s, the shoreline bandwidth density is (128 * 9.6)/5.8 Gb/s/mm = 26.5 GB/s/mm and the areal bandwidth density is (128 * 9.6)/(5.8 * 1.75) Gb/s/mm$^2$ = 15.1 GB/s/mm$^2$. For LPDDR6 at 12.8 GT/s, we scale the frequency: the shoreline bandwidth density is 35.3 GB/s/mm; areal bandwidth density is 20.2 GB/s/mm$^2$. Power efficiency for LPDDR6 is assumed to be the same as LPDDR5 at 2.8 pJ/b.

HBM4 designs are expected to use a bump pitch of 45u to 55u with 8 mm (die edge) x 2.5 mm (depth) at 6.4 GT/s[13,28]. Thus, the shoreline bandwidth density is 204.8 GB/s/mm and areal bandwidth density is 81.9 GB/s/mm$^2$. The power efficiency of 0.9 pJ/b from our HBM3 implementation will be used even for HBM4, even though it may be optimistic given the larger fan-out with twice the number of data lanes.

UCIe-S has an area of 1.143mm (die-edge) x 1.54mm (depth) at 110 um bump-pitch for a x32 link; UCIe-A has a fixed die-edge of 388.8 um across all bump-pitches with the depth of 1585, 1043, and 388 um for 55 um, 45 um, and 25 um bump-pitches respectively[23,25,26,29,30]. The *BW Density of UCIe* is calculated as follows: A doubly stacked UCIe-S at 32G has a b/w = 2 directions x 32 data lanes x 32 GT/s = 256 GB/s, bandwidth density is 224 GB/s/mm (linear) and 145.44 GB/s/mm$^2$ at 110 um bump-pitch. UCIe-A delivers 512 GB/s bandwidth for 64 data lanes; at 55um bump-pitch, the bandwidth density is 658.44 GB/s/mm and 416.27 GB/s/mm$^2$. Asymmetric UCIe has the same bandwidth density as bumps are the same for Transmitter vs Receiver.

For power consumption, when lanes grouped by their function (i.e., DQ + Wr Mask, Cmd, CRC, independently in each direction, in the asymmetric enhanced UCIe or the data lanes in the symmetric UCIe in each direction) are not in use, the dynamic power savings results in consuming $p$ fraction (($p$ = 0.15) of peak power (*Power Efficiency of UCIe* = 0.25 $to$ 0.5 $pJ/b$, for UCIe − A/S), as described in Table 1.

LPDDR6 protocol mapping on Asymmetric Enhanced UCIe: The calculation here is done for 74-Lane module in Fig. 4b with double stacking. Here command lanes are not the bottleneck since they match the maximum data transfer. The granularity of transfer in LPDDR6 is 288 bits: 256 bits of data + 32 bits of meta data (including ECC) with the x12 device arrangement[27].

Time needed (in UI) for:

$xR: x.\frac{576}{36} = 16x; \quad yW: y.\frac{576}{24} = 24y$ (1)

$xRyW: t_{xRyW} = \max(16x, 24y) = 8.\max(2x, 3y)$ (2)

Bandwidth Efficiency and Density is calculated as:

$BW_{eff} = \frac{(x+y).512}{74.t_{xRyW}} = \frac{(x+y).512}{74.8.\max(2x,3y)} = \frac{32.(x+y)}{37.\max(2x,3y)}$ (3)

$BW_{\{Density\}}(xRyW) = \frac{32.(x+y)}{37.\max(2x,3y)}.(BW\ Density\ of\ UCIe)$ (4)

Data power ratio is calculated as the ratio of power consumed for actual cache line data transfer in both directions over the total power consumption. For the SoC to Mem (S2M) direction, the 24 data plus 2 Wr mask are used for the data transfer for $24y$ UI and the remaining time is idle. $xRyW$ needs $96.(x + y)$ command bits. Hence, power consumed over $t_{xRyW}$:

$P_{S2M\_DQ\_WMask} = 26.(24y + (t_{xRyW} - 24y).p)$ (5)

$P_{S2M\_CMD} = 96.(x + y) + (10.t_{xRyW} - 96.(x + y)).p$ (6)

$$P_{S2M\_CRC} = \max(24y, 9.6(x+y)) \cdot (1-p) + 8 \cdot \max(2x, 3y) \cdot p \quad (7)$$
$$P_{M2S\_Data\_CRC} = 37 \cdot (16x \cdot (1-p) + 8 \cdot \max(2x, 3y) \cdot p) \quad (8)$$
$$P_{data} = \frac{512 \cdot (x+y)}{P_{S2M\_DQ\_WMask} + P_{S2M\_CMD} + P_{S2M\_CRC} + P_{M2S\_Data\_CRC}} \quad (9)$$
$$Power\ Efficiency\ for\ xRyW = \frac{Power\ Efficiency\ of\ UCIe}{P_{data}} \quad (10)$$

HBM3/4 protocol mapping on Asymmetric Enhanced UCIe: We use the 138-Lane UCIe as the basic building block for this analysis. Since the steps of this analysis is like that of LPDDR6 mapping, we omit the steps here due to page limits.

CXL.Mem on Symmetric UCIe: Each slot can carry two response headers in the Memory->SoC (M2S) direction and one request header in SoC->Memory (S2M) direction and each cache line data transfer requires 4 slots[31]. Thus, for $xRyW$, we get the following:

No. of Slots in S2M direction:
$$Slots_{S2M} = x + 5y \quad (11)$$
No. of Slots in M2S direction:
$$Slots_{Mem2SoC} = \frac{x+y}{2} + 4x = \frac{9x+y}{2} \quad (12)$$
$$Slots_{max} = \max(Slots_{Mem2SoC}, Slots_{SoC2Mem}) \quad (13)$$
$$BW_{eff} = \left(\frac{15}{16}\right) \cdot \frac{4(x+y)}{2 \cdot Slots_{max}} \quad (14)$$
$$BW\ Density\ for\ xRyW = \left(\frac{15}{16}\right) \cdot \frac{4(x+y)}{2 \cdot Slots_{max}} \cdot (BW\ Density\ of\ UCIe) \quad (15)$$
$$P_{data} = \left(\frac{15}{16}\right) \cdot \frac{4(x+y)}{(Slots_{Mem2SoC} + Slots_{SoC2Mem} + ((2 \cdot Slots_{max} - Slots_{SoC2Mem} - Slots_{Mem2SoC2}) \cdot p))} \quad (16)$$
$$Power\ Efficiency\ for\ xRyW = \frac{Power\ Efficiency\ of\ UCIe}{P_{data}} \quad (17)$$

CXL.Mem with Optimization on Symmetric UCIe: In S2M direction, we need $x + y$ slots (HS or G) for request headers for every $4y$ G-slots for data. Thus, if $x + y \leq \frac{4y}{15}$ all headers fit in the HS slots; else we will need $\frac{4y}{15} - (x + y)$ G-slots for the remaining headers. This can be expressed as:
$$Slots_{S2M} = \left(\frac{16}{15} * 4y\right) + \max\left(\left((x+y) - \frac{4y}{15}\right), 0\right) \quad (17)$$
Similarly, in the M2S direction, we need $\frac{x+y}{4}$ slots (HS or G) for the responses (since 4 responses can fit in one slot) for every $4x$ G-slots for data. Thus, if $\frac{x+y}{4} \leq \frac{4x}{15}$ all headers fit in the HS slots; else we will need $\frac{4x}{15} - \frac{x+y}{4}$ G-slots for the remaining headers. This can be expressed as:
$$Slots_{M2S} = \left(\frac{16}{15} * 4x\right) + \max\left(\left(\frac{(x+y)}{4} - \frac{4x}{15}\right), 0\right) \quad (18)$$
$$Slots_{max} = \max(Slots_{M2S}, Slots_{S2M}) \quad (19)$$
$$BW_{eff} = \frac{4(x+y)}{2 \cdot Slots_{max}} \quad (20)$$
$$BW\ Density\ for\ xRyW = \frac{4(x+y)}{2 \cdot Slots_{max}} \cdot BW\ Density\ of\ UCIe \quad (21)$$
Like the symmetric CXL.mem without optimization, with the exception that no slot is lost for CRC/ FEC/ Flit Hdr/ Credit, we calculate the data power ratio and realizable power efficiency with this approach can be expressed as:
$$P_{data} = \frac{4(x+y)}{(Slots_{Mem2SoC} + Slots_{SoC2Mem} + ((2 \cdot Slots_{max} - Slots_{SoC2Mem} - Slots_{Mem2SoC2}) \cdot p))} \quad (22)$$
$$Power\ Efficiency\ for\ xRyW = \frac{Power\ Efficiency\ of\ UCIe}{P_{data}} \quad (23)$$

### C. Results

Our proposed approaches outperform existing approaches using HBM4 and LPDDR6, across all metrics, as demonstrated in Figure 10, Figure 11, and Figure 12.

Our approaches with UCIe-A substantially outperform HBM4 with the same bump-pitch (55u), across all three metrics shown in Figure 10 and Figure 11. This demonstrates the advantage of UCIe-A over HBM4 for bandwidth density (linear and areal) and power efficiency. As expected, LPDDR6 being a 2D interconnect, performs substantially worse than even HBM4, but is provided here for completeness.

CHI does not perform as well as our other two approaches using CXL.Mem since its granules (equivalent of slots) are 20B (vs 16B for CXL) and there are less granules available for memory traffic. With memory-specific optimizations to CHI protocol mapped over UCIe, we expect it to perform better. Implementations with CHI in their internal fabric can easily convert to the optimized CXL.Mem and easily obtain the same performance benefits, if they want to.

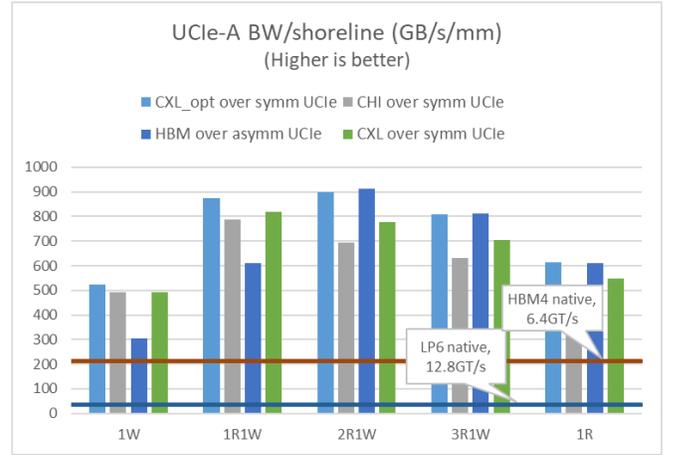

(a) Linear Bandwidth Density

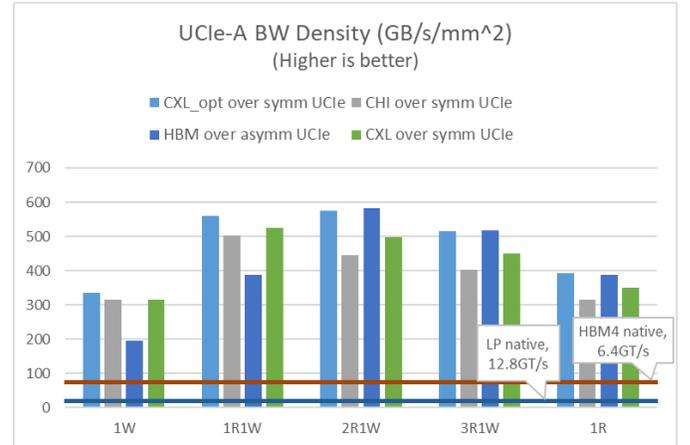

(b) Areal Bandwidth Density

Figure 10: Bandwidth Density Comparison of proposed approaches with UCIe-A vs existing LPDDR6 and HBM4

The power efficiency for LPDDR6 (at 2.8 pJ/b) is not provided here since an external interconnect will underperform on-package interconnects. Though with on-package optimizations LPDDR6 can potentially improve towards

HBM4. CXL.Mem over UCIe-A symmetric with optimization offers the best bandwidth density; achieving 6-10% improvement over CXL.Mem (without optimization) due to the additional G-slot (14 -> 15) and having twice the number of response headers in any slot. The power efficiency of our approaches with UCIe-S outperform HBM4 across all workloads, despite higher frequency and longer reach, due to its superior metrics (Table 1). Our proposed approach of mapping LPDDR6 and HBM4 over asymmetric UCIe perform slightly better than the optimized CXL.Mem on symmetric UCIe-A on quite a few traffic mixes (especially as read percentage increases) due to the fine grain power savings across Command vs Data in each direction; symmetric UCIe-A/-S can turn off all lanes or none.

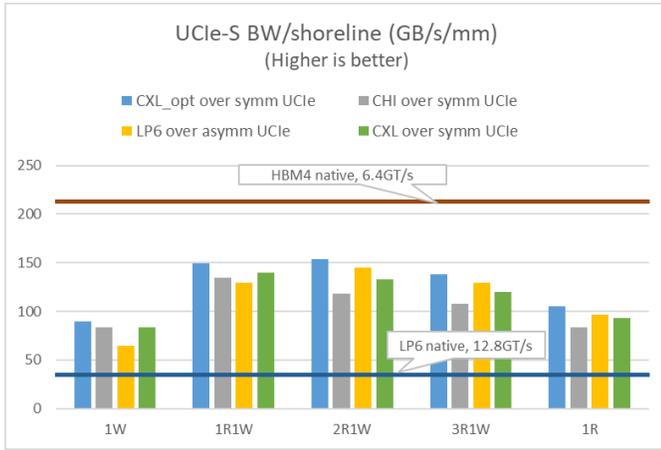

(a) Linear Bandwidth Density

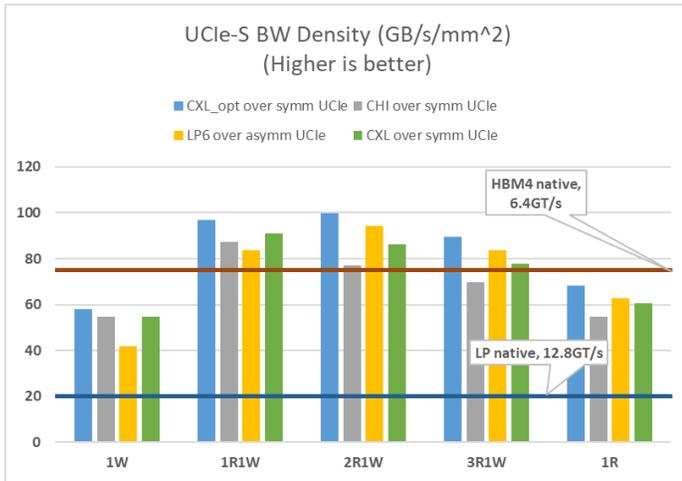

(b) Areal Bandwidth Density

Figure 11: Bandwidth Density Comparison of proposed approaches with UCIe-S vs existing LPDDR6 and HBM4

Our approaches with UCIe-S at 110u bump pitch outperform LPDDR6 across all metrics and traffic mixes (Figure 11 and Figure 12). Optimized CXL.Mem on symmetric UCIe-S performs best. Interestingly, our approach with UCIe-S substantially outperforms HBM4 for most workloads, even with cheaper packaging. Additionally, cheaper LPDDR6 dies using wire bonding with our approach will significantly reduce cost and power over HBM4. HBM4 does offer better bandwidth per shoreline (Figure 11a) since it goes deeper into die (2.8mm vs UCIe-S's 1.5mm). UCIe-S can easily double the bandwidth per shoreline stacking twice the number of modules since it can handle a channel reach of 25 mm, still leaving an ample 22mm for routing (vs HBM4 routing is less than 2mm) or by doubling the frequency since it has enough headroom to double the data rate. The power efficiency of our approaches with UCIe-S comes close to HBM4 across all workloads (e.g., 10-20% for optimized CXL on symmetric UCIe), despite higher frequency (32G vs 6.4G) and longer reach (25mm vs 2mm).

Based on the results, we establish that CXL.Mem with optimization on symmetric UCIe offers the best power-efficient performance with the additional benefit of leveraging the existing IPs and infrastructure the ecosystem has developed.

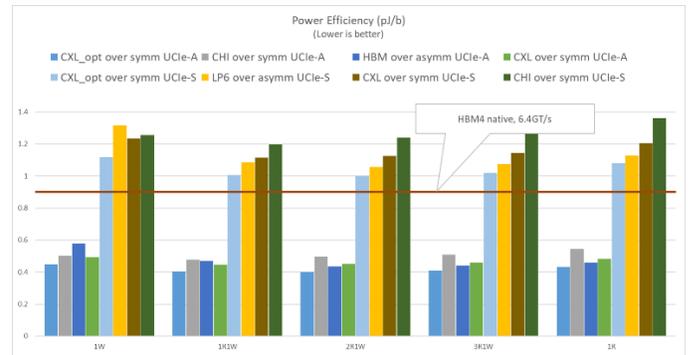

Figure 12: Power Efficiency comparison of proposed approaches based on UCIe-A and UCIe-S against existing HBM4 (LPDDR6)

## V. Conclusions

Our on-package memory approaches deliver superior power-efficient performance over existing approaches. To improve bandwidth density further, UCIe should increase the operating frequency while continuing to be bump-limited with constant power efficiency. Further bandwidth density improvement can be achieved in planar interconnects by stacking multiple UCIe ports on the same shoreline, at the expense of increased routing layers on the package, which is a reasonable trade-off. Additionally, since UCIe bandwidth scales with bump pitch reduction in advanced packaging due to increased bump density, our proposed approaches will continue to meet the future memory bandwidth demands. Advances in packaging technology will mitigate the capacity challenges by adding more memory devices on-package, taking advantage of the high bandwidth density of our proposed approaches. To realize the power-efficient and cost-effective benefits of our proposed approaches, we need to standardize these approaches and work on ecosystem enabling and adoption, both from the SoC and memory vendors.

## References

1. D. Das Sharma and R. Mahajan, "Advanced packaging of chiplets for future computing needs'', invited comment article, Nature Electronics, July 2024 (DOI:

# Appendix

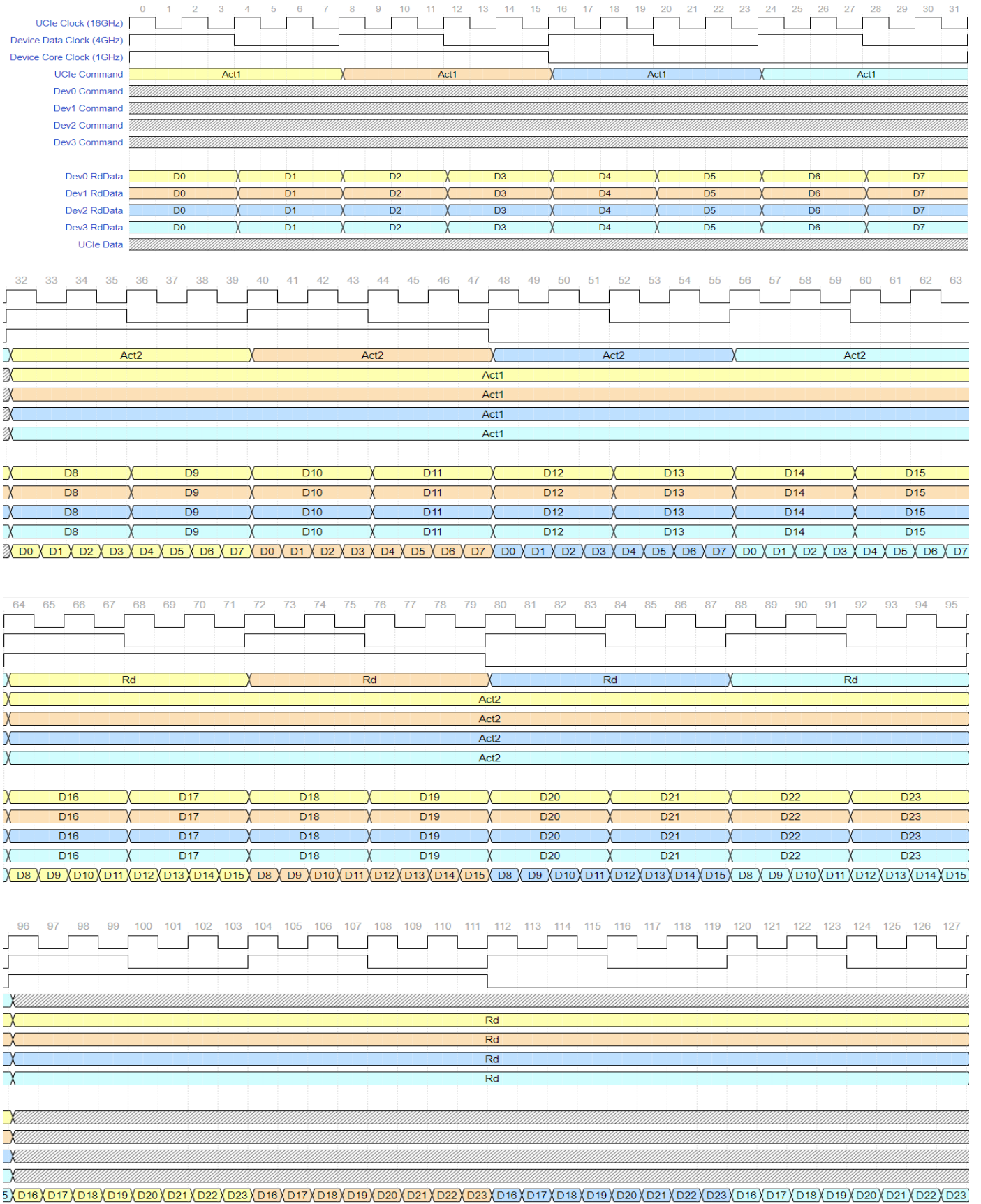

*Figure 13: Time multiplexing example for Reads at 8-bit granularity showing pipelining of Activate and Read Commands. Four LPDDR6 devices are aggregated behind the logic die. Each color represents the command or data for a different x12 LPDDR6 device with a burst length of 24. Each of the 4 stacked sub-figures is showing 32 clocks of the 16GHz clock which is used for the 32GT/s data rate over UCIe. Each sub-figure is a continuation in time, as indicated by the cycle number, from the previous one.*